\documentclass[12pt,a4paper]{article}
\usepackage{amsmath,amsthm,amssymb,epsfig,graphicx}
\usepackage[section] {placeins}
\usepackage{float}
\usepackage{booktabs}
\usepackage{perpage}
\usepackage{mathptmx}
\usepackage{CJKutf8}
\usepackage{graphicx}
\usepackage{subfig}
\usepackage[T1]{fontenc}
\MakeSorted{figure}
\MakeSorted{table}
\usepackage{setspace}
\begin{document}

%\begin{frontmatter}

\title{New Immersed Boundary Method with Irrotational Discrete Delta Vector for Droplet Simulations with Large Density ratio}

\author{Chia Rui Ong and Hiroaki Miura}
\maketitle
%\address{Department of Earth and Planetary Science, 
%Graduate School of Science, The University of Tokyo}

\begin{abstract}
The Immersed Boundary Method (IBM) is one of the popular one-fluid mixed Eulerian-Lagrangian 
methods to simulate motion of droplets. While the treatment of a moving complex 
boundary is an extremely time consuming and formidable task in a 
traditional boundary-fitted fluid solver, the one-fluid methods provide a relatively easier way to 
track moving interfaces on a fixed Cartesian grid since the regeneration of a 
mesh system that conforms to the interface at every time step can be avoided. In the IBM, a series 
of connected Lagrangian markers are used to represent a fluid-fluid interface and the boundary 
condition is enforced by adding a forcing term to the Navier-Stokes equations. To enable this, a 
discrete delta function is necessary for interpolation of velocity field and extrapolation of immersed 
boundary forcing between Eulerian grids and Lagrangian markers.  \par
It is known that the IBM suffers two problems. One is spontaneous generation of unphysical kinetic 
energy, which is known as parasitic currents and also appear in other one-fluid methods as well, and 
the other is spurious reconstruction of interface. These two problems need to be solved for useful 
long-time-scale simulations of droplets with high density ratio and large surface tension. This 
work detects that the discrete delta function is the cause of unphysical parasitic currents. 
Specifically, the irrotational condition is not preserved when the common discrete delta function is 
used to spread the surface tension from Lagrangian markers to Cartesian grid cells. To solve this problem, 
a new scheme that preserves the irrotational condition is proposed to remove the spurious 
currents. Furthermore, for a smooth reconstruction of an interface, a 
B-spline fitting by least squares is adopted to relocate the Lagrangian markers. The 
conventional and new interpolation schemes are implemented in a multigrid finite volume Direct 
Numerical Simulation (DNS) solver and are subjected to standard test cases. It is confirmed that the 
unphysical parasitic currents are substantially reduced and droplet's surface fluctuation is eliminated in the 
new scheme. The new scheme is also applied for simulations of axisymmetric free-fall droplet. In this test, the numerical results 
coincide well with experimental data. \par
\end{abstract}

%\begin{keyword}
%Immersed boundary method \sep Parasitic currents \sep 
%Discrete delta function \sep Direct numerical simulation 
%\sep Incompressible flow
%\end{keyword}

\newpage
\tableofcontents %Table of contents
\cleardoublepage %The first chapter should start on an odd page.

%\end{frontmatter}

\newcommand{\ddel}[2]%new command with 2 arguments
		   {\frac{\partial#1}{\partial#2}}
\newcommand{\ddelE}[1]{\frac{\partial}{\partial#1}}
\newcommand{\ddeld}[2]{\frac{\partial^{2}#1}{\partial^{2}#2}}
\newcommand{\ddelc}[3]{\frac{\partial^{2} #1}{\partial#2\partial#3}}
\newcommand{\cov}[1]{\vec{u} \cdot  \bigtriangledown{#1}}
\newcommand{\deld}[1]{\bigtriangledown \cdot #1}
\newcommand{\curl}[1]{\bigtriangledown \times #1}
\newcommand{\der}[2]{\frac{d#1}{d#2}}
\newcommand{\dert}[2]{\frac{d^{2}#1}{d#2^{2}}}
			
\newpage

\section{Introduction}
\label{sec:intro}
The Direct Numerical Simulation (DNS) of multiphase flow is a challenging problem. The difficulties 
lie in the non-linearity of the interfacial boundary conditions and the correct representation of 
the interface that is freely transported by fluids. 
There are generally two types of methods to tackle these difficulties according to how the location 
of an interface is tracked down: the interface-fitted method and the one-fluid Eulerian-Lagrangian 
models (in short form as one-fluid model). In the interface-fitted model 
\cite{Chen2017}\cite{Gros2018}\cite{ScarZa1999}\cite{Zheng2016}, a grid is regenerated at 
every time step so that the interfacial grid cells constantly conform to the interfacial boundary. 
The grid regeneration process is highly computational demanding. Furthermore, it has difficulties in 
dealing with the separation and merging of two interfaces. Despite these disadvantages, it has advantages that the interface 
can be accurately represented and discretization of the interfacial boundary condition is straigtforward. 
On the other hand, the one-fluid models have been becoming popular in recent decades because the Navier-Stokes equations are 
solved on the Cartesian grid. The interface is tracked down either by some explicit functions, which 
are transported by the advection equation, or Lagrangian markers. This interface tracking allows a substantial reduction of computational costs. The discretization of equations is much more 
straightforward than the curvilinear mesh used in the interface-fitted model. Although the time 
consuming process of grid regeneration at every time step can be avoided, the adaptive mesh is 
sometimes needed for better accuracy. \par

One-fluid models can be further categorized into several methods according to the way to ensure 
that the interfacial boundary conditions are satisfied and interfaces are tracked. There are two schemes 
to treat the interfacial boundary conditions: 1) the sharp interface method 
and the 2) continuous interface method. In the sharp interface method, the interfacial boundary 
conditions are explicitly incorporated into the difference equations when discretizing the 
Navier-Stokes equations. Examples of the sharp interface method are the Immersed Interface 
Method (IIM) \cite{LeeLe03} and the ghost-fluid method \cite{Fed99}. In the continuous interface 
method proposed by Brackbill et al \cite{Brack92}, a smooth forcing term which has a support of 
several grid spacings is added into the Navier-Stokes equations and the force is spread to the 
Cartesian grid by an extrapolation scheme. A significant advantage of the continuous interface 
method against the sharp interface method is that it is straightforward to implement it in an 
existing fluid solver without changing discretization schemes. Meanwhile, a disadvantage is that a 
force term needs to be extrapolated to the Cartesian cell centers or faces by a regularized discrete 
delta function \cite{Bao2016}\cite{LiuMori}\cite{Yang2009}. This reduces the accuracy of the model down to the first order due to the smearing of the interface \cite{BeLe92}\cite{Mori2007}\cite{Tornberg2004}. \par

Depending on the way of tracking interfaces, one-fluid models can be categorized into three main 
groups: (1) The Immersed Boundary Method (IBM) \cite{peso}\cite{Peskin2002}\cite{Mittal}\cite{Esma04}\cite{Juric96} in which 
interfaces are represented by a series of connected or disconnected Lagrangian markers; (2) The 
Volume-Of-Fluid (VOF) \cite{Hi_Ni}\cite{No_Wo}\cite{You82} in which a volume function is used to represent the volume fraction of a specific fluid in a grid cell; (3) The Level-Set (LS) \cite{Os_Se}\cite{Suss94}\cite{Suss_Puc} in which a smooth 
level set function is used to represent the location of an interface. There are other one-fluid 
models such as the diffuse interface method \cite{Jacq1999} in which a vastly different strategy is 
adopted. In the diffuse interface method, the density of fluid is determined by the generalized 
diffusion equation. \par

In this work, we focus on the implementation of the IBM with the continuous interface method for 
simulations of a droplet. We believe that there are three important conditions that need to be 
satisfied for a successful long-time-scale simulation of a droplet or droplets with moderate-to-large 
density ratio between two fluids. First, the force spreading operator should be irrotational to avoid 
unphysical spurious flow parallel to the interface. Second, the sum of surface tension force over the 
computational domain should be as small as possible (ideally zero). Third, a stable reconstruction of the interface for a smooth 
distribution of curvature. However, we found that the IBM often fails to satisfy one or more 
conditions above. It will be shown in this work that failing to satisfy the first condition leads to 
the generation of the so-called parasitic currents. In other words, to eliminate the parasitic 
currents, the curl of discrete delta vector interpolated onto the Cartesian grid from the Lagrangian 
markers has to be zero strictly. Previous works have already discussed this problem 
in different one-fluid models (see \cite{Lafau} and \cite{Yokoi13}), and some 
solutions have been proposed. For example, Francois et al \cite{Frans06} have devised a 
balanced-force algorithm, which produces an exact balance between the pressure gradient force and the 
surface tension force, to eliminate it in the context of VOF. Jamet et al \cite{Jamet02} used the diffuse interface method with 
an energy conserving discretization to remove it. In this work, we proposed a new approach that uses 
an irrotational extrapolation scheme based on the discrete gradient of the Heaviside function in the IBM formulation. \par

In the IBM, the curvature of a surface at each marker can be approximated by the absolute value of difference of tangent vectors of adjacent vectors over the distance between them. As the distance between two adjacent markers is variable, the total surface tension force may not be zero. Moreover, the markers often move in a haphazard way and the curvature is very sensitive to the locations of markers due to the property of interpolation spline curve. This leads to unphysical flow fields near the boundary when the curvature fluctuates heavily. As a result, the latter two conditions are often not satisfied and stability is severely impeded. In order to solve these problems, we decided to adopt the global B-spline curve to interpolate and redistribute the markers evenly at every time step.  \par

The new model proposed above satisfies the first and third conditions while the condition two is only approximately satisfied, because the new spreading method may not conserve the total surface tension force. In the droplet simulations, this is usually a minor issue compared to the instability caused by the parasitic currents and curvature fluctuation. To demonstrate the validity of our new approach, test cases have been done. The test cases are static and free oscillating droplet in both two dimensional and cylindrical axisymmetric domains. A further set of free-fall droplet simulations were performed to demonstrate its capability. \par

The overview of this paper is as follows. Section 2 is devoted to the formulation and discretization 
schemes of the finite volume solver and the immersed boundary method. Details of the problem of parasitic currents and implementation of our new scheme are given in section 3. The numerical results are presented in section 4. The conclusion is given in the final section. \par

\newpage
\section{Continuum formulation and numerical discretization}
\subsection{Flow solver}
\begin{figure}[h]
  \centering
  \includegraphics[scale=0.6]{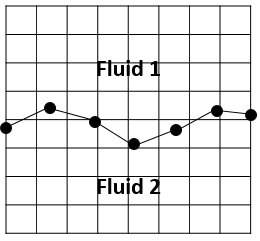}
  \caption{Fluid 1 and fluid 2 are separated by an interface which is represented by a series 
		       of connected Lagrangian markers in the IBM formulation}
  \label{pp_ibm}
\end{figure}

\begin{figure}[h]
  \centering
  \includegraphics[scale=0.6]{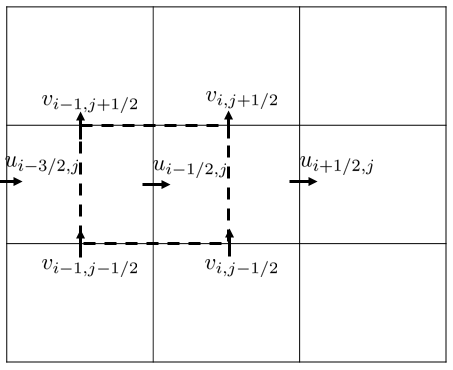}
  \caption{Definition of indices used in the discretization of Navier-Stokes equations. The staggering grid is adopted in which velocity vectors are defined on cell surfaces and scalars such as pressure and density are defined on cell centers.}
  \label{cpdomain}
\end{figure}

The formulation of the Navier-Stokes equations for incompressible flow with the immersed boundary 
(IB) forcing term is: 
\begin{equation}
  \label{eqn:ns_1}
   \rho \ddel{\vec{u}}{t} + \rho \deld{(\vec{u} \vec{u})}  = -\bigtriangledown{p} + 
   \deld{ (\mu \nabla \vec{u})}+ \vec{f_{ib}}, \; and
\end{equation}
\begin{equation}
  \label{eqn:divg}
  \deld{\vec{u}} = 0.
\end{equation}
Equations \eqref{eqn:ns_1} and \eqref{eqn:divg} are the momentum equation and continuity 
equation respectively where $\vec{u}$ is velocity, $\rho$ is density, $p$ is pressure, and $\mu$ 
is kinematic viscosity. The forcing term $\vec{f_{ib}}$ in equation \eqref{eqn:ns_1} is added to 
satisfy the interfacial boundary conditions and its specific form will be defined in the next 
section. In the IBM, the simple staggering Cartesian mesh system is used in the finite-volume fluid 
solver, while the fluid-fluid interface is represented by a series of connected Lagrangian markers 
(Figure \ref{pp_ibm}). We employ the fractional step method \cite{Cho} to solve equations above. In the fractional step method, the Navier-Stokes equations are solved in two steps by splitting the momentum equation into two parts: 
\begin{equation}
  \label{eqn:ns1}
   \rho^{n+1} \frac{\vec{u}^* - \vec{u^n}}{\Delta t}  = RHS + \vec{f_{ib}}, \; and
\end{equation}
\begin{equation}
  \label{eqn:ns2}
   \rho^{n+1} \frac{\vec{u}^{n+1} - \vec{u}^*}{\Delta t}  = -\bigtriangledown{p^{n+1}}.
\end{equation}
where $u^*$ is the intermediate velocity, which does not necessarily satisfy the divergence-free 
condition, $\Delta t$ is the time step, $RHS$ is the advective and diffusive terms, and the 
superscript $n$ means scalar or vector values at the $n^{th}$ time step. The density is determined from 
the indicator function which will be given in the next section. The temporal discretization methods 
for the advection term and the viscous stress term are the Adams-Bashforth scheme and the 
Crank-Nicholson scheme, respectively. The spatial discretization of these two terms about a velocity 
component $u$ for the control volume centered at face $(i-1/2,j)$ as shown in Figure \ref{cpdomain} 
is given by (superscript is ignored):
\begin{equation}
\label{eqn:ad_bf}
  \begin{split}
  \oint \rho \deld{(\vec{u} u)} \; dV &= \sum\limits_{all faces} \rho u \vec{u} \cdot d\vec{s} \\
  &= \rho_{i,j} \left ( \frac{u_{i+1/2,j} + u_{i-1/2,j}}{2} \right )^2 \Delta y_{i-1/2,j} - 
     \rho_{i-1,j} \left ( \frac{u_{i-1/2,j} + u_{i-3/2,j}}{2} \right )^2 \Delta y_{i-1/2,j} \\
     &\quad + \bar{\rho}_{i-1/2,j+1/2} \left ( \frac{v_{i,j+1/2} + v_{i-1,j+1/2}}{2} \right ) 
       \left ( \frac{u_{i-1/2,j} + u_{i-1/2,j+1}}{2} \right ) \Delta x_{i-1/2,j} \\
     &\quad - \bar{\rho}_{i-1/2,j-1/2} \left ( \frac{v_{i,j-1/2} + v_{i-1,j-1/2}}{2} \right ) 
       \left ( \frac{u_{i-1/2,j} + u_{i-1/2,j-1}}{2} \right ) \Delta x_{i-1/2,j} \\
  \end{split}
\end{equation}
\begin{equation}
\label{eqn:cr_ni}
  \begin{split}
  \oint \deld{ (\mu \ \bigtriangledown u)} \; dV &= 
     \sum\limits_{all faces} \mu \bigtriangledown u \cdot d\vec{s} \\
  &= \mu_{i,j} \frac{u_{i+1/2,j} - u_{i-1/2,j}}{\Delta x_{i-1/2,j}}\Delta y_{i-1/2,j} - 
     \mu_{i-1,j} \frac{u_{i-1/2,j} - u_{i-3/2,j}}{\Delta x_{i-1/2,j}}\Delta y_{i-1/2,j} \\
  &\quad + \mu_{i-1/2,j+1/2} \frac{u_{i-1/2,j+1} - u_{i-1/2,j}}{\Delta x_{i-1/2,j}}\Delta y_{i-1/2,j} \\
	&\quad - \mu_{i-1/2,j-1/2} \frac{u_{i-1/2,j} - u_{i-1/2,j-1}}{\Delta x_{i-1/2,j}}\Delta y_{i-1/2,j} \\
  \end{split}
\end{equation}
where $\Delta x_{i-1/2}$ and $\Delta y_{i-1/2}$ are the local horizontal and vertical grid spacings. 
$d\vec{s}$ is the normalized normal vector of control volume's surface vector and 
$\bar{\rho}_{i-1/2,j+1/2}$ denotes the averaged value of density on grid node $(i-1/2,j+1/2)$ 
(similarly for viscosity coefficient). \par
By taking the divergence of equation \eqref{eqn:ns2} and enforcing the 
divergence of velocity field at all time steps is zero (equation \eqref{eqn:divg}), we get:
\begin{equation}
  \label{eqn:ps}
   \frac{\deld{\vec{u^*}}}{\Delta t} = \deld{ \left( \frac{1}{\rho^{n+1}}\bigtriangledown p^{n+1} \right) }
\end{equation}
This is a Poisson's equation. The Laplacian term is discretized by the second-order centered difference 
scheme. It is then solved by the multigrid method \cite{Uday}\cite{Briggs2000}. Finally, 
velocity field at the next time step can be obtained by substituting the updated pressure 
back into equation \eqref{eqn:ns2}. \par
The same discretization schemes are adopted for the discretization of the Navier-Stokes equations in 
the axisymmetric cylindrical coordinate system. \par

\subsection{Immersed boundary method}
The basic formulation of the IBM in this work largely follows Unverdi and Tryggvason 
\cite{Unverdi_trygg92}. In the IBM, the fluid-fluid interface is represented by a series of 
Lagrangian markers that are allowed to freely move on the fixed Cartesian mesh. To exchange 
information between the interface and the fixed mesh, interpolation and extrapolation are thus 
required. This is done by the discrete version of delta function. The discrete delta function is in 
the same form for both interpolation and extrapolation and it is defined by:
\begin{equation}
  \label{eqn:delta}
  \delta(\vec{x} - \vec{X}) = \frac{1}{\Delta x \Delta y} \phi \left( \frac{dx}{\Delta x}\right) 
                              \phi \left( \frac{dy}{\Delta y}\right)
\end{equation}
where $\vec{X}$ is the location of a marker, $\vec{x}$ is an arbitrary location at which the value of 
delta function is sought, $\Delta x$ and $\Delta y$ are the local $x$ and $y$ grid spacings 
respectively, dx and dy are $x$ and $y$ components of vector $\vec{x} - \vec{X}$, and $\phi$ is a continuous function with compact support. The grid spacing (in both $x$ and $y$ directions) is assumed to be constant in the vicinity of the fluid-fluid interface in this work. The function $\phi$ has to satisfy several important properties. For example, the sum over all grid points must be equal to one. Further details are given in \cite{Peskin2002}. There are many candidates for $\phi$ \cite{LiuMori}\cite{Bao2016}. In this work, we use the first-order Chebyshev polynomial and its expression is:
\begin{equation}
  \label{eqn:cherby}
  \phi(r) = 
  \begin{cases}
    \frac{1}{4\Delta s} \left( 1 + \cos{\frac{\pi r}{2\Delta s}} \right) & \text{if } 0 \leq |r| < 
    2\Delta s\\
    0 & \text{if } 2\Delta s \leq |r|\\
  \end{cases}
\end{equation}

Using equation \eqref{eqn:delta}, the surface tension force on Lagrangian markers can be spread onto the Eulerian grid and velocity field can be interpolated back into the Lagrangian markers. \par

As the sharp interface between two fluids is replaced by a smooth differentiable delta function that spreads from two to four grid spacings, the physical properties such as the density and viscosity inside the smooth transition region become statistical. In other words, physical properties no longer follow thermodynamic or physical law in the transition region, in contrast to the diffuse interface method in which the profiles of physical properties are derived based on the second law of thermodynamics \cite{Jam1995}\cite{Jacq1999}. It is assumed that the distributions of the physical properties in the smearing zone follow the shape of smooth Heaviside step function, i.e. 
\begin{equation}
  \label{eqn:indicator}
  \psi = \psi_1 + (\psi_2 - \psi_1)I,
\end{equation}
where $I$ is the indicator function and $\psi$ represents a physical variable and the subscript 
denotes fluid 1 or 2 (Figure \ref{pp_ibm}). The value of $\psi$ ranges from zero to one, with being zero in fluid 1 and one in fluid 2. The indicator function satisfies the following Poisson's equation:
\begin{equation}
  \label{eqn:ind_psn}
  \deld{\bigtriangledown{I}} = \deld{\vec{\delta}}
\end{equation}

The immersed boundary forcing term $f_{ib}$ has to be defined from the jump conditions across the 
interface. Assuming that the surface tension coefficient is constant, the interfacial jump conditions 
with no mass transfer are:
\begin{equation}
  \label{eqn:interface_bc1}
  \hat{n} \times (\vec{u}_2 - \vec{u}_1) = 0
\end{equation}
\begin{equation}
  \label{eqn:interface_bc2}
  \rho_1(\vec{u}_1 - \vec{u}_i) \cdot \hat{n} = \rho_2(\vec{u}_2 - \vec{u}_i) \cdot \hat{n} = 0
\end{equation}
\begin{equation}
  \label{eqn:interface_bc3}
  p_1 - \tau_1 \cdot \hat{n} = p_2 - \tau_2 \cdot \hat{n} - \nabla_s \sigma + \sigma \kappa \hat{n}
\end{equation}
where the subscripts $1$, $2$, and $i$ denote fluid $1$, fluid $2$, and the interface respectively, 
$\rho$ is the density, $p$ is the pressure, $\vec{u}$ is the velocity, $\tau$ is the viscous stress, 
$\nabla_S$ denotes the surface gradient operator, $\sigma$ is the surface tension constant, $\kappa$ 
is the curvature, and $\hat{n}$ is the normalized vector normal to the surface. \par
The first and second equations above simply mean that the velocity is continuous across the interface. 
The third equation can be rewritten as
\begin{equation}
  \label{eqn:jump_cond}
  [\Pi]_{1,2} = \sigma \kappa \hat{n}
\end{equation}
where we have lumped the pressure and viscous stress jumps to the left hand side of the 
equation and denote it by $[\Pi]_{1,2}$. The square bracket means a discontinuity or jump across the 
interface. The right hand side is surface tension force. In the continuous forcing IBM, the 
interface between two fluids is smeared out and thus the physical properties such as density and 
viscosity are assumed to follow Heaviside distribution. 
Therefore, the bracket in equation \eqref{eqn:jump_cond} is to be replaced by the directional 
derivative of Heaviside function and a corresponding delta function is added to the RHS as follows
\begin{equation}
  \label{eqn:jump_condf}
  \bigtriangledown \Pi = \sigma \kappa \delta \hat{n}
\end{equation}

From this form, it is found that the surface tension force can be interpreted as a flux equivalent to 
the combination of pressure flux and viscous stress. Thus it is physically reasonable to add an extra 
surface tension forcing term to the Navier-Stokes equations to enforce the interfacial jump conditions 
between two fluids. However, it is found that the surface tension force becomes rotational even when the curvature and surface tension coefficient is constant if the discrete delta function \eqref{eqn:delta} is substituted into equation \eqref{eqn:jump_condf}, leading to a  generation of parasitic currents. \par

The other interfacial boundary condition of continuous velocity can be satisfied by using the discrete delta function (eq \eqref{eqn:delta}) to interpolate velocity from the Eulerian grid onto Lagrangian markers because the velocity is defined on markers. 

\begin{equation}
  \label{eqn:jump_condu}
  \vec{U}(\vec{X}) = \Sigma_{i,j} \vec{u}(\vec{x}) \delta(\vec{x} - \vec{X}) \Delta x \Delta y
\end{equation}
where $\vec{U}(\vec{X})$ is the velocity on markers, $\vec{u}(\vec{x})$ is velocity on Eulerian 
grid, and the summation is over the entire computational domain. However in one-fluid finite volume methods, 
the numerical computations are executed in the discrete Cartesian space and each variable is averaged 
over the size of a grid cell. After the independent variables are updated to the next time step, the 
exact location of the interface is lost because the mapping between the Lagrangian markers and 
volume-averaged variables in the Cartesian grid is not one-to-one. Therefore, the location of the 
interface can only be updated approximately. This may lead to numerical error, or noise, in the 
locations of Lagrangian markers. 
Moreover, discretization errors inevitably lead to some small noise in the interpolated velocity on the Lagrangian markers and subsequently the small noise results in small-scale surface fluctuation when the locations of markers are updated. Furthermore, it is well known that a high-order direct polynomial fitting is very sensitive to data points. These factors give rise to the large fluctuation of curvature if smoothing process is not performed. Eventually the fluctuation leads to unphysical pressure gradient and movements of droplets because the surface tension force is a linear function of the curvature. \par
\begin{figure}[h]
  \centering
	\begin{tabular}{c}
	\subfloat[]{
  \includegraphics[scale=0.4]{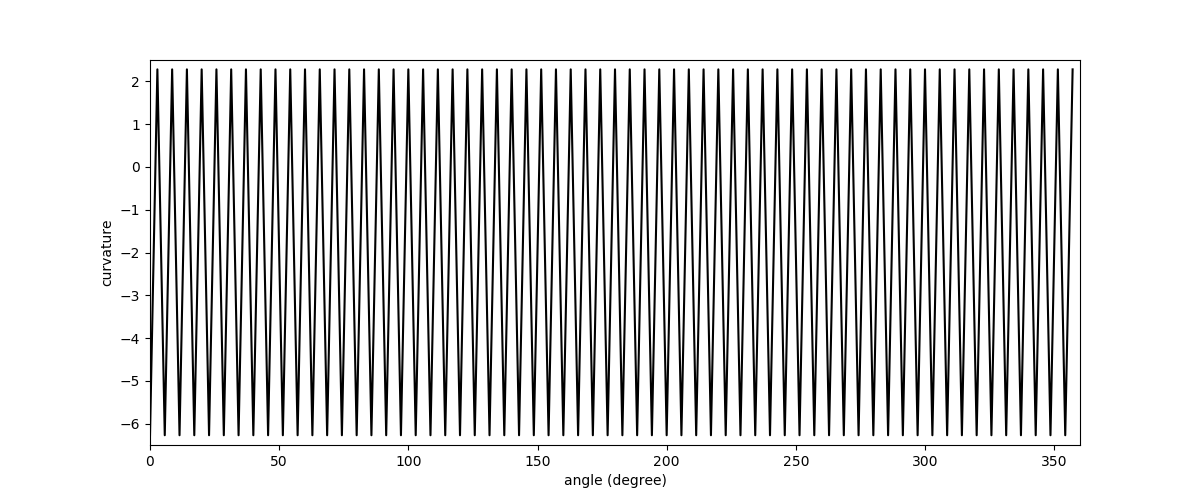}} \\
	\subfloat[]{
  \includegraphics[scale=0.4]{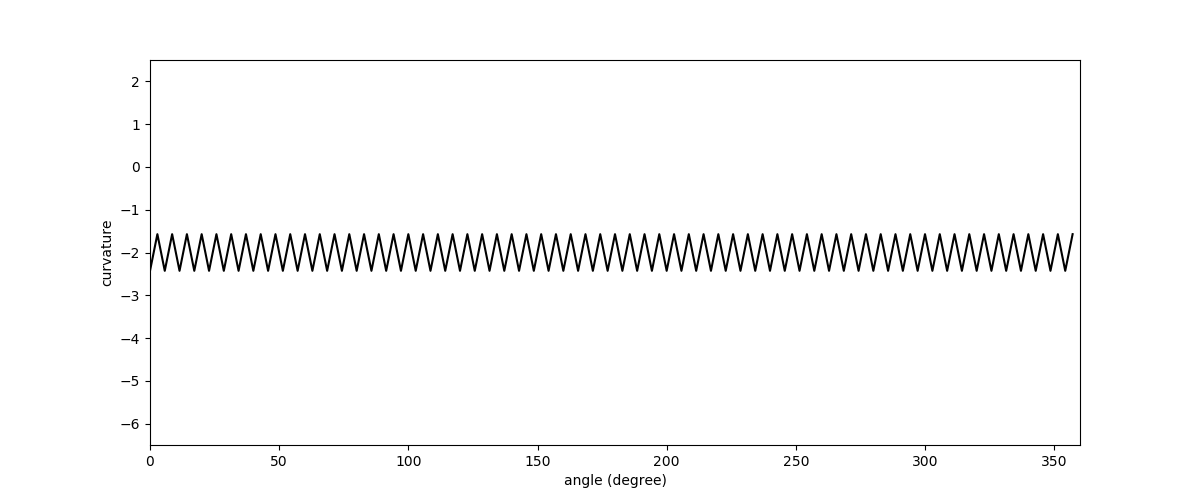}} \\
	\end{tabular}
  \caption{Plot of curvature at each marker when $g$ in eq. \eqref{eqn:interpol_cir} is (a) $0.001$ and (b) $0.0001$.}
  \label{interpol_1}
\end{figure}

We consider a perturbed circular interface with coordinates of markers $(x_i,y_i)$ given by the following equations to illustrate the sensitivity of the polynomial fitting to the locations of markers. 
\begin{equation}
\begin{split}
\label{eqn:interpol_cir}
x_i &= 0.5 + 0.5(1 \pm g)sin\left(\frac{2\pi i}{n}\right), \\
y_i &= 0.5 + 0.5(1 \pm g)cos\left(\frac{2\pi i}{n}\right), \\
\end{split}
\end{equation}
where $g$ is the perturbation constant, $n$ is the total number of markers and $i$ is an integer from $0$ to $n-1$. The number of markers is $n=126$ and they are equally spaced along the interface. The plus or minus sign in eq. \eqref{eqn:interpol_cir} is taken according as $i$ is odd or even. Two plots of curvature against the angle are shown in figure \ref{interpol_1} for $g=0.001$ and $g=0.0001$ respectively. The amplitude of the fluctuation of curvature is larger than $4$ when $g=0.001$, which corresponds to just $0.1\%$ perturbation of the diameter. \par
\begin{figure}[h]
  \centering
  \includegraphics[scale=0.4]{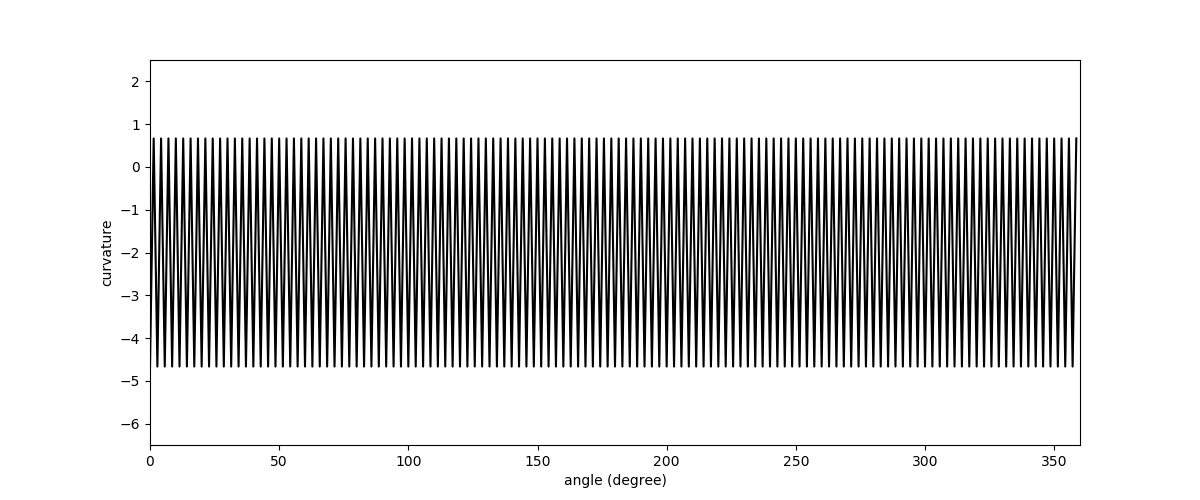}
  \caption{Plot of curvature at each marker when the distance between two neighboring markers is non-uniform.}
  \label{interpol_2}
\end{figure}

Besides numerical errors, addition or deletion of markers can cause fluctuation too. They are a necessary step because two adjacent markers may move too close or too far to each other as time evolves in actual simulations. In the work of \cite{Singh07}, when the distance between two adjacent markers is below half of the grid size, one of the two markers is deleted. If the distance is over 1.5 grid size, an additional marker is added at the midpoint of the two markers. To illustrate the fluctuation, we again consider the circular interface in which a marker is added in the midway between two neighboring markers in every other interval. The plot of curvature is shown in figure \ref{interpol_2}. Similar to the previous case, a strong fluctuation can be observed. \par 
To eliminate this instability problem, a smooth enough reconstruction of the interface is needed to suppress high frequency fluctuation of curvature. In the next two sections, we will explain how to define a new spreading operator to remove parasitic currents and employ the global B-spline fitting by least squares to enable smooth surface reconstruction in detail. \par

\subsection{Non-inertial frame}
When simulating a free fall droplet, it may take a long falling distance until it 
reaches its terminal velocity. In order to avoid preparing a large and dense 
computational domain, the non-inertial framework is adopted in the free-fall droplet tests. 
\begin{equation}
  \label{eqn:non-inertial}
  \ddel{\vec{u}}{t} + \vec{a}  = \vec{J}
\end{equation}
In equation \eqref{eqn:non-inertial}, an extra fictitious force $\vec{a}$ is added 
to the momentum equation to account for the non-inertial reference frame centered on the droplet. 
Following the work of Komrakova et al \cite{Komra}, the expression for $\vec{a}$ is
\begin{equation}
  \label{eqn:fictitious}
  a_y = c_1a_{y1} + c_2a_{y2},
\end{equation}
\begin{equation}
  \label{eqn:fic_1}
  a_{y1} = \frac{(COM_o - COM_n)}{\Delta t}, \; and
\end{equation}
\begin{equation}
  \label{eqn:fic_2}
  a_{y2} = \frac{(COM_n - COM_{n-1})}{\Delta t}. 
\end{equation}
where $COM_o$ is the initial center of mass of the droplet, $COM_n$ and $COM_{n-1}$ are center of mass 
at $n^{th}$  and $n-1^{th}$ time steps. The non-inertial force acts only in the vertical component of 
the momentum equation in the direction of the gravitational acceleration. The two constants $c_1$ and 
$c_2$ are both set to 0.1. Once the fictitious acceleration is calculated from the equations above, 
it is added as a source term into the Navier-Stokes equations when solving for intermediate velocity 
at every time step. \par

\subsection{Conservation of mass}
The advection of markers by the discrete delta function does not follow the conservation law on the 
Eulerian grid. Thus, it is not guaranteed that the mass of the dropelt is always conserved. Following 
the work of Udaykumar et al \cite{Uday97}, the area enclosed by the markers is calculated every 20 
time steps. The bisection method is then applied if the deviation from the original volume exceeds a 
discrepancy error of order $1 \times 10^{-6}$. The discrepancy error is defined as $|A_n - A_o|/A_o$, where 
$A_n$ is the area at $n^{th}$ time step and $A_o$ is the initial area. In other words, markers are 
moved inwards or outwards in the normal direction to the interface when the volume growth is 
overestimated or underestimated. \par

\section{Problem of parasitic currents}
\subsection{Rotational force spreading}
\begin{figure}[h]
  \centering
  \includegraphics[scale=0.6]{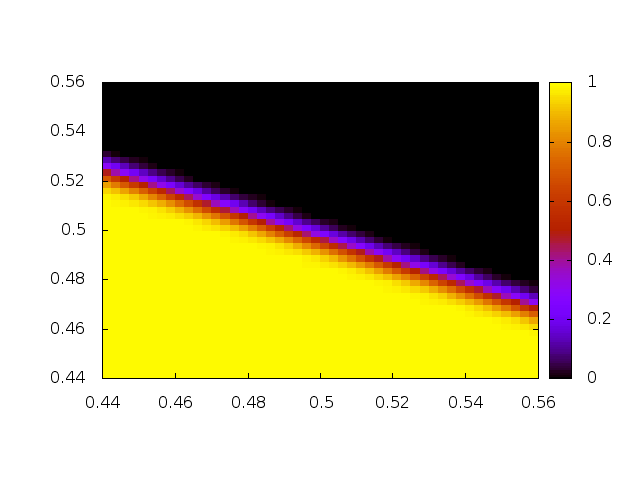}
  \caption{Plot of $P$, which is solution of the Laplacian equation $\nabla^2 P = \nabla \cdot \vec{F}$. A straight-line interface declined at an angle of $25^{o}$ exerts a constant forcing $\vec{F}$ downwards. The magnitude of force is one.}
  \label{Heaviside_phi}
\end{figure}

\begin{figure}[h]
  \centering
  \includegraphics[scale=0.6]{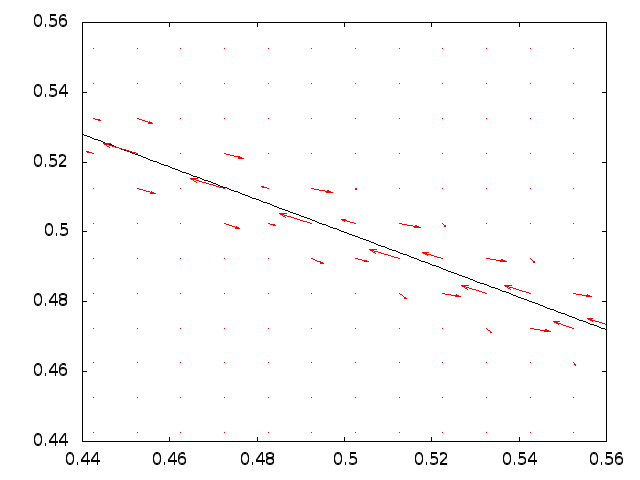}
  \caption{Velocity field $(u,v)=(P_{i,j} - P_{i-1,j} - F_{i-1/2,j},P_{i,j} - P_{i,j-1} - F_{i,j-1/2})$ generated by the discrepancy between the numerical solution of $\nabla^2 P = \nabla \cdot \vec{F}$ and the true solution. The black solid line is the interface.}
  \label{Heaviside_uv}
\end{figure}

\begin{figure}[h]
  \centering
  \includegraphics[scale=0.6]{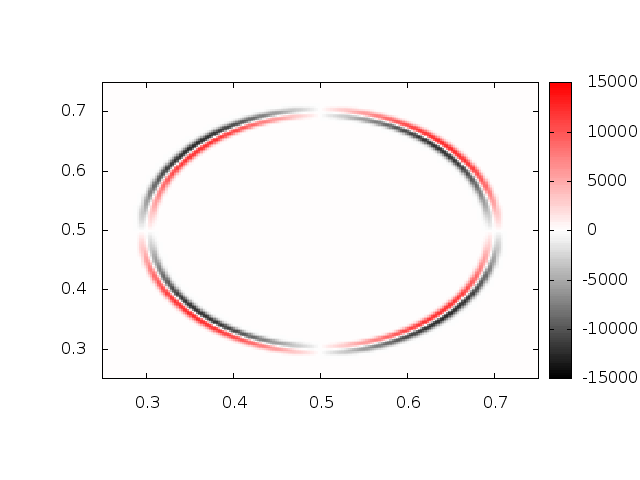}
  \caption{A plot of the curl of discrete delta vector for a circular interface of radius $0.2$ exerting a constant force of magnitude one inwards.}
  \label{Heaviside_curl}
\end{figure}

The problem of parasitic currents is commonly found in one-fluid multiphase flow methods 
\cite{Yokoi13}\cite{Lafau}. In the work of \cite{Lafau}, they showed by numerical experiments with the VOF 
that the maximum absolute value of fluid velocity caused by parasitic currents is proportional 
to $\sigma/\mu$. In order to see the cause of parasitic currents, let us define the spreading 
of a constant force, $\vec{f}$, by the discrete delta function as a vector function and call it discrete delta vector:
\begin{equation}
  \label{eqn:delta_vec}
  \vec{\delta}(\vec{x} - \vec{X}) = \vec{f} \delta(\vec{x} - \vec{X})
\end{equation}
where $\delta$ is given in \eqref{eqn:delta}. A continuous delta vector should be 
irrotational mathematically. This can be easily seen by writing the delta vector in terms of Heaviside function 
and note that curl of gradient is zero. 
\begin{equation}
  \label{eqn:delta_vec_hea}
  \nabla \times \vec{\delta} = \nabla \times \nabla H = 0
\end{equation}
In equation \eqref{eqn:delta}, the discrete delta vector is splitted into orthogonal directions following the mesh. 
However, this splitting causes the discrete delta vector fail to satisfy the irrotational property, 
resulting in the generation of parasitic currents. In order to demonstrate this problem, let us 
consider a simple linear differential problem in which a straight interface inclined at an angle with 
a constant forcing spread by the delta vector function defined by equation \eqref{eqn:delta_vec}. The 
magnitude of force is one and the force is pointing downwards into the lower region. In mathematical 
form: 
\begin{equation}
  \label{eqn:poisson_delta}
  \nabla^2 P = \nabla \cdot \vec{\delta} %\varrho
\end{equation}
where $P$ is a scalar function that resembles pressure in physcial sense when the force is surface tension force. 
The analytic solution is simply the Heaviside step function obtained by integrating equation 
\eqref{eqn:cherby}. After substituting equation \eqref{eqn:delta} into \eqref{eqn:poisson_delta} and 
solving it numerically by a standard Poisson solver, the results are shown in figure 
\ref{Heaviside_phi}. The grid size is $0.005$. The small discrepancy between the exact discrete delta 
forcing and the numerical solution, i.e. $\vec{\delta} - \nabla P$, causes the circulation along the 
interface (figure \ref{Heaviside_uv}). 
Figure \ref{Heaviside_curl} further shows a plot of curl for a circular interface. Since the curl of 
the discrete delta vector is not zero in this case again, the solution obtained from solving the 
Poisson's equation will not be a perfect Heaviside function.
The magnitude of curl is constant at a fixed shortest distance to the interface when the grid spacing 
decreases. And it can be shown that the maximum magnitude of curl is at the order of $\Delta x$ by 
brute force calculation. Thus, the numerical solution does converge to the analytic solution at first 
order accuracy. Note that this calculation is based on the particular case in which the discrete delta function $\phi$ is in the cosine form (eq \eqref{eqn:cherby}) and there is a possibility that another discrete delta function might not cause the same kind of problem. Therefore, searching for such delta function can be a way to solve this problem. In this work, however, we are taking another approach introducing a new discrete delta vector that satisfies the irrotational condition. \par
Having understood the cause of parasitic currents, we now propose a new discrete delta vector that is irrotational. \par

\subsection{Irrotational discrete delta vector}

\begin{figure}[h]
  \centering
  \includegraphics[scale=0.6]{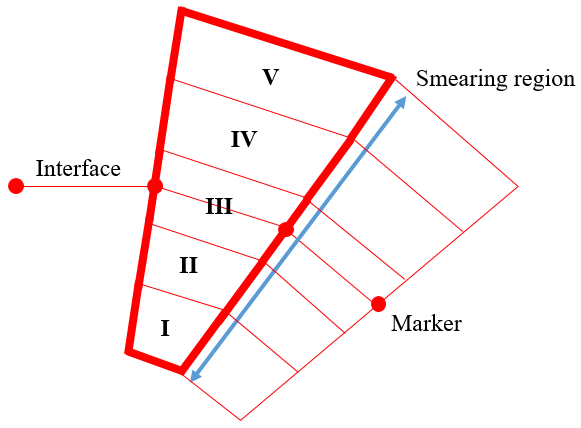}
  \caption{Schematic view of the construction of Heaviside function based on the 
	         discrete delta function (eq \eqref{eqn:cherby})}
  \label{pp_hf}
\end{figure}

\begin{figure}[h]
  \centering
  \includegraphics[scale=0.6]{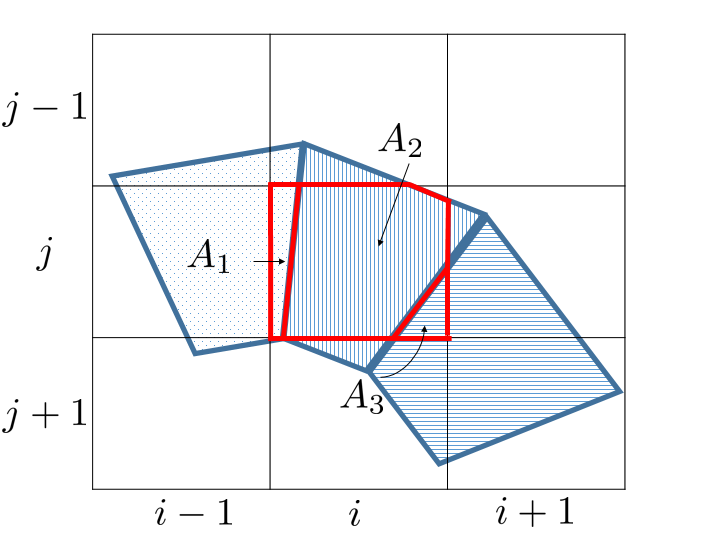}
  \caption{Three Polygons constructed along an interface. The surface tension force in each polygon is different. The value of Heaviside's function is obtained by adding the contributions from each overlapped polygon, i.e. $H_{i,j}=A_1 H_1 + A_2 H_2 + A_3 H_3$}
  \label{pp_df}
\end{figure}

An easy way to ensure that the discrete delta vector is irrotational is that we construct it from a discrete Heaviside function by defining the delta vector as
\begin{equation}
  \label{eqn:delta_hea}
  \vec{\delta} = \nabla H
\end{equation}
By discretizing the directional derivative, the discrete delta vector can be written as:
\begin{equation}
  \label{eqn:delta_hea_xy}
  \begin{split}
  \vec{\delta}_{i-1/2,j} &= f_{i-1/2,j}\delta_{i-1/2,j} = f_{i-1/2,j}(H_{i,j} - H_{i-1,j}) \\
  \vec{\delta}_{i,j-1/2} &= f_{i,j-1/2}\delta_{i,j-1/2} = f_{i,j-1/2}(H_{i,j} - H_{i,j-1}) \\
  \end{split}
\end{equation}
where $H_{i,j}$ is the discrete Heaviside function defined on the cell's center. $f_{i-1/2,j}$ is the surface tension and its calculation will be explained in the next section. In this way, the discrete delta vector is guarenteed to be irrotational, that is$(f_{i,j-1/2}\delta_{i,j-1/2} - f_{i-1,j-1/2}\delta_{i-1,j-1/2})\Delta x + (f_{i-1/2,j-1}\delta_{i-1/2,j-1} - f_{i-1/2,j}\delta_{i-1/2,j})\Delta y = 0$, if the surface tension force is constant. Since the interface is represented by a series of lines and markers, we can easily construct a discrete Heaviside function based on the linear configuration of the interface. As shown in 
figure \ref{pp_hf}, a polygon area bounded by thick line about each segment along the interface is 
constructed. Inside each polygon, the area is divided into five regions I, II, III, IV, and V. The 
magnitude of discrete Heaviside function of each region is computed based on the integration of 
the discrete delta function (equation \eqref{eqn:cherby}) along the direction normal to the interface: 
\begin{equation}
  \label{eqn:new_hea}
  \begin{split}
  H &= \int_{-2\Delta s}^{r_i} \frac{1}{4 \Delta s} \left( 1 + \cos{\frac{\pi r}{2\Delta s}} \right) \; dr \\
    &= \frac{1}{4 \Delta s}\left( r_i + \frac{2\Delta s}{\pi} sin \left( \frac{\pi r_i}{2\Delta s} + \frac{1}{2\Delta s}
   \right) \right) \\
  \end{split}
\end{equation}
$\Delta s$ is a constant because the grid size is constant in the vicinity of the interface, $r_i$ is the distance to the interface, and the interface is assumed to be located at $r=0$. Hence, by substituting values of $r_i$ according to regions I, II, III, IV, and V into equation \eqref{eqn:new_hea}, we obtain
\begin{equation}
  \label{eqn:new_hea2}
  H = 
  \begin{cases}
    0 & \text{Region I } \\
    0.409 & \text{Region II }\\
    0.5 & \text{Region III }\\
    0.909 & \text{Region IV }\\
    1 & \text{Region V }\\
  \end{cases}
\end{equation}
Suppose we want to find the discrete Heaviside function $H_{i,j}$ on a particular cell $(i,j)$ as shown in figure \ref{pp_df}. There are three piece-wise Heaviside's polygons that overlap with cell $(i,j)$. Each overlapped region contributes to $H_{i,j}$ by the amount of $A_I H_I$, where $A_I$ is the overlapped area and $H_I$ is the corresponding piece-wise Heaviside function. So $H_{i,j}$ inside cell $(i,j)$ is then obtained by adding them. \par

\begin{figure}[h]
  \centering
  \includegraphics[scale=0.5]{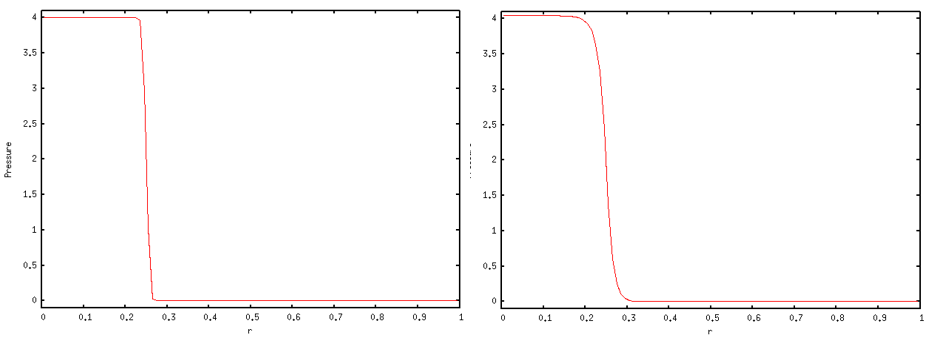}
  \caption{Left: pressure plot with timestep = 0.0001. Right: pressure plot with timestep = 0.005}
  \label{psi_com}
\end{figure}

In the traditional continuous forcing approach, the surface tension force is added to equation 
\eqref{eqn:ns1} when the Navier-Stokes equations are solved by the fractional step method, but 
this causes unnecessarily larger smearing of forcing distribution. To demonstrate this, we have 
applied the IBM to incompressible 1D Navier-Stokes equations in cylindrical coordinates. 
We have run two simulations of timesteps $0.005$ and $0.0001$. The pressure distributions after one time step are shown in figure \ref{psi_com}. 
The effective range of surface tension force becomes larger as the time step gets larger 
which apparently extends outside of the original range of the discrete delta function. The 
solution will still converge to the true solution in the limit of $\Delta t \to 0$. 
The excessive spreading is due to that the immerbsed boundary force is added into equation \eqref{eqn:ns1}, and thus the range of spreading becomes proportional to the time step. Therefore, we apply the surface tension force in the projection step, i.e. equation \eqref{eqn:ns2} instead of \eqref{eqn:ns1}, in order to avoid this unnecessary larger smearing of the force. \par
\begin{equation}
  \label{eqn:ns1_new}
   \rho^{n+1} \frac{\vec{u}^* - \vec{u^n}}{\Delta t}  = RHS, \; and
\end{equation}
\begin{equation}
  \label{eqn:ns2_new}
   \rho^{n+1} \frac{\vec{u}^{n+1} - \vec{u}^*}{\Delta t}  = -\bigtriangledown{p^{n+1}} + \vec{f_{ib}}.
\end{equation}

\section{B-spline reconstruction}
Surface tension is a linear function of local curvature. Accurate computation of distribution of 
curvature along the interface is thus important. As it has been shown that a direct polynomial interpolation of 
Lagrangian markers is very sensitive to their locations, we propose to use the method of global 
B-spline curve fitting by least squares \cite{deBoor}. 
A B-spline curve is controlled by control points and its knot vector. A knot vector is a non-decreasing sequence, denoted by $(t_{-d},t_{-d+1},...,t_{g+d},t_{g+d+1})$. Each knot, $t_i$, represents a data point in the parametric space. The symbol $d$ denotes the degree of interpolating polynomial and $g$ denotes the number of segments. Assume that a fluid-fluid interface is comprised of $n$ connected Lagrangian markers. We define the $i^{th}$ knot, $t_i$, as the distance travelled along the interface (which is a set of connected piece-wise line segments) from the first Lagrangian marker to the $i^{th}$ marker. The first marker is arbitrarily set. Based on the argument above, the B-spline curve is given by:
\begin{equation}
  \label{eqn:bspline}
  \begin{split}
  B_{i,d}(x) &= \frac{x - t_i}{t_{i+d} - t_i}B_{i,d-1} + \frac{t_{i+d+1} - x}{t_{i+d+1} - t_{i+1}}B_{i+1,d-1}\\
  B_{i,0}(x) &= 1 \\
  \end{split}
\end{equation}
$B_{i,d}$ is a polynomial of $x$ of degree $d$ and non-zero only in range between knot $t_i$ and knot $t_{i+d+1}$. A global interpolating B-spline polynomial $B(x,t,d)$ is 
\begin{equation}
  \label{eqn:bcurve}
  B(x,t,d) = \Sigma_{i=-d}^{g}c_i B_{i,d}(x)
\end{equation}
The coefficients $c_i$ are determined by the least squares method. In the least squares method, the following sum of squares is minimized to determine all the coefficients $c_i$:
\begin{equation}
  \label{eqn:bspline_ls}
  \begin{split}
  \Delta = \Sigma_{r=1}^{n}\left[ y_r - \Sigma_{i=-d}^{g}c_i B_{i,d}(x_r) \right]^2
  \end{split}
\end{equation}
The quality of the B-spline curve is often dependent on the arrangement of knots and its degree. 
However, the determination of knots' locations to best-fit the data points is often difficult and 
depends on the nature of the data itself. Thus, throughout this work, we employ a simple strategy to 
determine the knot sequence. The knot sequence is simply set to an arithmetric progression with the 
common distance equal to $1.5\Delta s$ where $\Delta s$ is the grid spacing. The order of B-spline 
is $3$ to ensure the continuity of curvature. This fitting process by least squares is performed at 
every time step to redistribute the Lagrangian markers so that the distance between two adjacent 
markers is always a constant. The curvature is defined as the rate of change of tangent vector 
along the interface. In discrete form, it can be written as:
\begin{equation}
  \label{eqn:ten_f}
  \kappa_i = |\frac{\vec{T}_{i-2\Delta s} - \vec{T}_{i+2\Delta s}}{\Delta s}|,
\end{equation}
where $\vec{T}_{i-2\Delta s}$ is the tangent vector whose location is at $i^{th}$ 
marker and ${i+1}^{th}$ marker. A tangent vector at any point on the curve can be obtained by directly differentiating the B-spline curve as follows:
\begin{equation}
  \label{eqn:b_tan}
  \vec{T} = (\der{B(x,t,d)}{x},\der{B(y,t,d)}{y})/|L|
\end{equation}
where $|L|$ is the length of the vector $(\ddel{B(x,t,d)}{x},\ddel{B(y,t,d)}{y})$. \par
In the axisymmetric case, the curvature is given by the following equation, which can be derived by applying the variational principle on the Gibb's free energy, 
\begin{equation}
  \label{eqn:curv_ax}
  \kappa = \left[ \frac{\der{y}{s}}{x\sqrt{1+\left(\der{y}{s}\right)^2}} + 
	         \frac{\dert{y}{s}}{\sqrt[3]{1+\left(\der{y}{s}\right)^2}}\right]
\end{equation}

In this work, this step is no longer needed because all markers are redistributed by the smooth global B-spline curve. However, it may be very difficult to directly extend the current method to a closed 2D surface embedded in 3D domain. Instead of global fitting, one may try local fitting by least-squares to reconstruct a smooth interface iteratively. We leave this examination as a future task. \par

\newpage
\section{Results}
The numerical results are presented to elucidate the differences between new and 
conventional methods in this section. Static droplet and oscillating droplet simulations, which are 
the standard test cases for multiphase flow simulations \cite{Torres}\cite{Fyfe}, have been done to 
demonstrate the validity of the new method. Numerical simulations of free-fall droplet will be given 
in the last part of this section to show its robustness in long-time-scale simulations. \par

\subsection{Static droplet simulation}
\label{sec:ibm_ver}
A static droplet simulation is a very simple test case in which a circular or spherical droplet is 
placed in the computational domain without initial motion to verify that the surface tension force is 
properly distributed onto the Cartesian grid, creating a pressure jump across the interface. This test 
has been doen in both 2D and axisymmetric domains. The pressure jump can be obtained analytically 
from the Young-Laplace's equation:
\begin{equation}
  \label{eqn:lap_rel}
  \Delta P = \sigma \left( \frac{1}{R_1} + \frac{1}{R_2} \right)
\end{equation}
where $R_1$ and $R_2$ are the principal radii of curvature. \par

\begin{figure}[h]
  \centering
  \includegraphics[scale=0.6]{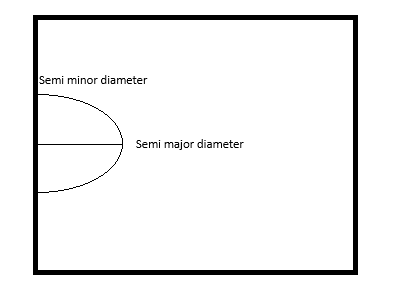}
  \caption{Schematic view of the initial configuration of axis-symmetric droplet simulations and the definition of semi minor and major diameter.}
  \label{ocd_ell}
\end{figure}

In our 2D static droplet simulations, a droplet of diameter $0.4$ is placed in the center of the 
computational domain. This setup is the same for the axisymmetric case, except that the droplet is 
placed at the center of left boundary of the domain (Figure \ref{ocd_ell}). All simulations in the 
static droplet tests were run only for one time step of $1 \times 10^{-3}$ to observe the pressure 
difference between the ambient fluid and the droplet. The magnitude of surface tension coefficient 
$\sigma$ is set to one. The exact pressure jump is $5$ for a 2D droplet and $10$ for an 
axisymmetric droplet. Numerically, we computed the total pressure jump by taking the difference of 
the averages of pressure inside and outside the droplet excluding the interfacial boundary cells:
\begin{equation}
  \label{eqn:sd_pres_diff}
  \Delta P = \frac{1}{N_i} \sum_{N_i} P_i - \frac{1}{N_a} \sum_{N_a} P_a
\end{equation}
where $N_i$ and $N_a$ are the number of cells of droplet's fluid and ambient fluid, $P_i$ and $P_a$ 
are the pressure inside and outside the interface respectively. \par
\begin{figure}[h]
  \centering
	\subfloat[][]{%{0.5/textwidth}
    \includegraphics[scale=0.35]{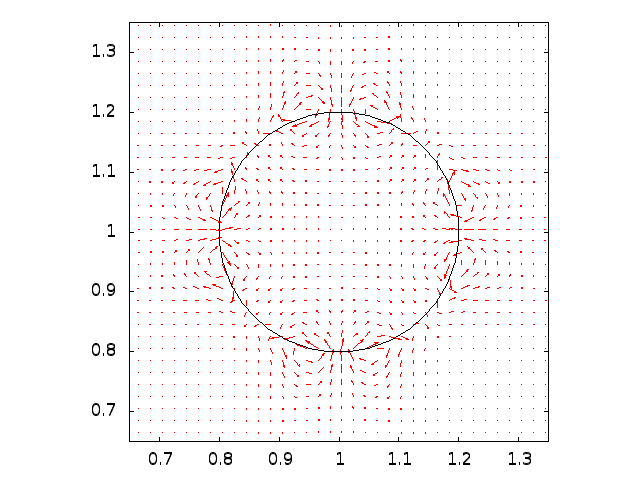}}
	\subfloat[][]{%{0.5/textwidth}
    \includegraphics[scale=0.35]{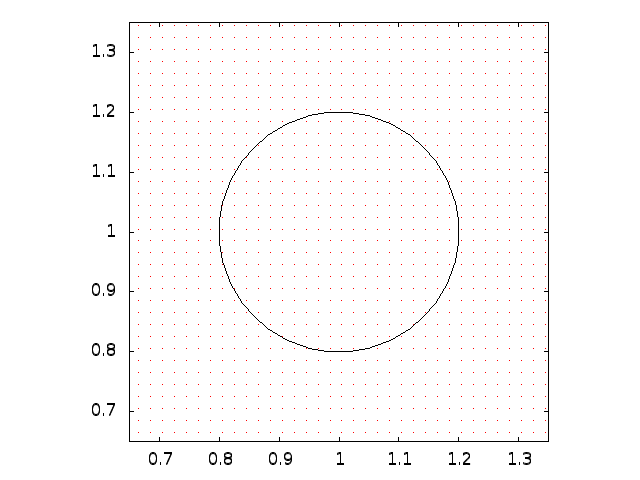}}
  \caption{Velocity field of the 2D static droplet simulation after the steady state solution is reached. a) conventional scheme b) new scheme.}
	%, 2D. c.) old scheme, axis-symmetry. d.) new scheme, axis-symmetry.}
  \label{staticdrop_pc}
\end{figure}

Tables \ref{table:staticdrop_2d} and \ref{table:staticdrop_axis} give the error of pressure jump 
for different grid sizes. The results are rounded to 5 significant numbers. The density of the ambient fluid is $0.01$ and the droplet's density 
is $1$. The discrepancy of pressure jump from the correct value is around $0.2\%$ for the conventional scheme which is 
mainly caused by the conventional discrete delta vector (equation \eqref{eqn:delta}) that does not preserve 
the irrotational property. The discrepancy is generally smaller for the new scheme and it is mainly 
due to small fluctuation of the curvature computed from the 
least squares B-spline interpolation. Moreover, the pressure difference converges properly to the analytic value. 
As shown in Figure \ref{staticdrop_pc}, the parasitic currents 
circulating along the interface in the conventional scheme was successfully eliminated when the new 
scheme is used. In conclusion, the results show that the pressure jump of a static droplet is 
accurately reconstructed by the new scheme and its magnitude converge properly to the analytic value 
with increasing grid resolution. This indicates that the interfacial boundary conditions are satisfied 
in the solutions. \par

\begin{table}
\begin{center}
  \begin{tabular}{|c|c|c|c|c|}
   \hline
   \hline
   $\Delta x$ & $||p - p_{true}||_{L1}$ (conventional scheme) &  $||p - p_{true}||_{L1}$ (new scheme) \\
   \hline
   $0.02$ & $1.6475e-3$ & $1.1082e-3$ \\
   \hline
   $0.01$ & $1.4186e-3$ & $4.1935e-4$ \\
   \hline
   $0.005$ & $1.1080e-3$ & $5.5086e-4$ \\
   \hline
   $0.0025$ & $1.0822e-3$ & $2.2961e-4$ \\
   \hline
   $0.00125$ & $1.0744e-3$ & $1.2822e-4$ \\
   \hline
  \end{tabular}
  \caption{The error of pressure difference of a 2D circular droplet for different grid spacings with density ratio $100$}
  \label{table:staticdrop_2d}
\end{center}
\end{table}

\begin{table}
\begin{center}
  \begin{tabular}{|c|c|c|}
   \hline
   $\Delta x$ & $||p - p_{true}||_{L1}$ (conventional scheme) &  $||p - p_{true}||_{L1}$ (new scheme)\\
   \hline
   $0.02$ & $5.0019e-3$ & $4.0890e-3$ \\
   \hline
   $0.01$ & $8.8084e-4$ &  $1.3275e-3$ \\
   \hline
   $0.005$ & $8.6586e-4$  &  $1.2220e-3$ \\
   \hline
   $0.0025$ & $1.8752e-3$  & $4.8848e-4$ \\
   \hline
   $0.00125$ & $2.3319e-3$  & $2.2793e-4$ \\
   \hline
  \end{tabular}
  \caption{The pressure difference of an axisymmetric spherical droplet for  different grid spacings with density ratio $100$}
  \label{table:staticdrop_axis}
\end{center}
\end{table}

\subsection{Oscillating droplet simulation}
The purpose of simulations of an oscillating droplet in this subsection is to test the performance of 
the new scheme when dynamic movement of markers is included. A circular droplet is slightly perturbed 
to an elliptic shape at the beginning, and then it undergoes a free oscillation. The locus of the 
interface is given by:
\begin{equation}
  \label{eqn:osc}
  \begin{split}
  x &= x' + r cos(\theta)\beta \\
  y &= y' + r sin(\theta)/\beta \\
  \end{split}
\end{equation}
where $n$ is $2$, $R$ is $0.2$, and $r$ is the droplet's radius. $x'$ and $y'$ are the coordinates 
of the center of droplet. They are $(1,1)$ and $(0,1)$ respectively for the 2D droplet and the 
axisymmetric droplet simulations. The size of the computational domain is $2$ by $2$. The droplet's 
density and viscosity are $0.01$ and $0.01$ while the ambient fluid's density and viscosity are $0.01$ and 
$0.00001$. The surface tension coefficient is one. The initial magnitude of perturbation $\beta$ is 
set to $1.03$. Due to unbalanced distribution of pressure jump along the interface, the droplet will 
start to restore back to a circle by pushing inwards on the side of larger curvature and outwards on 
that of smaller curvature. For a small preturbation, the oscillation period can be predicted 
analytically. The period of a 2D oscillating droplet is given by the following 
equation (see \cite{Fyfe}): 
\begin{equation}
  \label{eqn:osc_p}
  T = 2\pi\sqrt{\frac{(\rho_{droplet} + \rho_{ambient}) r^3}{(n^3 - n)\sigma}}
\end{equation}
where $\rho_{droplet}$ is droplet's density and $\rho_{ambient}$ is ambient fluid's density. 
In the axisymmetric case, the period is given by (for derivation, see \cite{hydro}):
\begin{equation}
  \label{eqn:aosc_p}
   T = 2\pi\sqrt{\frac{((n+1)\rho_{droplet} + n \rho_{ambient})r^3}{n(n-1)(n+1)(n+2)\sigma}} 
\end{equation}
Substituting n by $2$, the periods are $0.23057$ and $0.19935$ respectively. \par

\begin{figure}[h]
  \centering
	\subfloat[][]{%{0.5/textwidth}
    \includegraphics[scale=0.35]{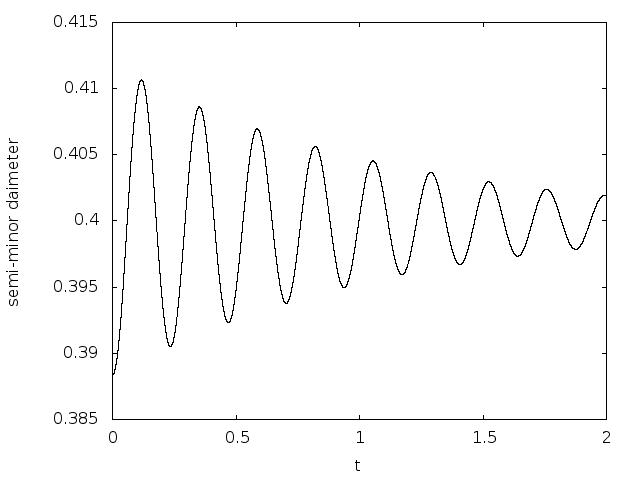}}
  %\end{subfloat}%
	\subfloat[][]{%{0.5/textwidth}
    \includegraphics[scale=0.35]{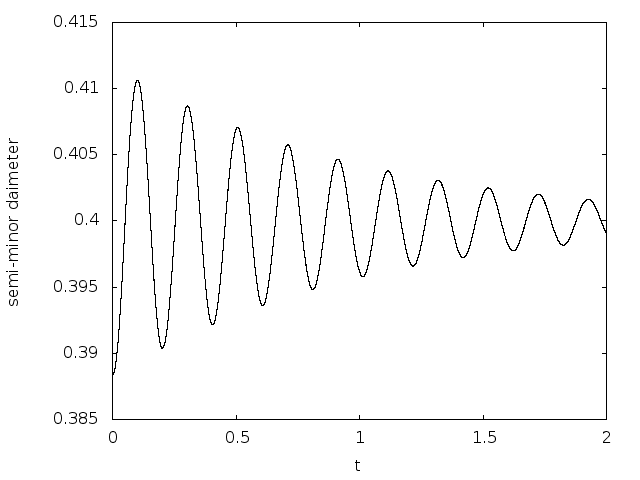}}
  %\end{subfloat}%
	\caption{Plot of semi minor diameter of an oscillating droplet against time at grid size $0.01$ using our new scheme. a) 2D. b) Axisymmetry.}
  \label{ocd_dia}
\end{figure}

Three simulations were performed for different grid resolutions to verify the convergence of the new 
scheme. The grid sizes are $0.02$, $0.01$, and $0.005$ respectively. In other words, there are 20, 
40, and 80 cells across the equivalent diameter of the droplet. The first oscillation cycle is 
observed and compared to the analytic solutions. The results for different grid sizes are summarized 
in Table \ref{table:osc}. The results show that the new scheme does converge properly to the analytic 
solution as resolution increases. The convergence rate is approximatetly at the first order. Note that the 
error is slightly smaller for the conventional scheme. This may be due to the large parasitic 
currents enhancing the frequency by increasing the restoration energy. There is possibly a second reason due to the smoother surface reconstructed by B-spline least squares fitting. The time evolution of the semi minor diameter for simulations of grid size $0.01$ using the new scheme is shown in figure \ref{ocd_dia}. \par

\begin{figure}[h]
  \centering
	\subfloat[][]{%{0.5/textwidth}
    \includegraphics[scale=0.35]{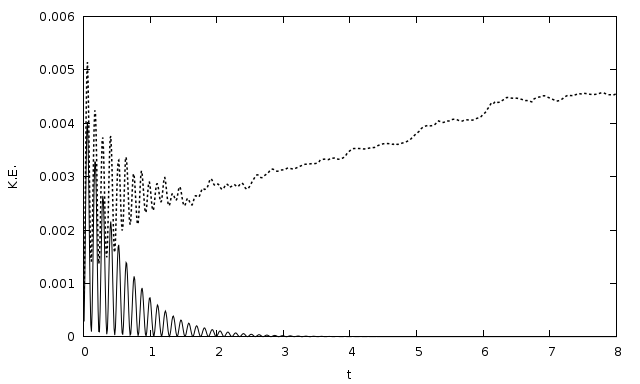}}
  %\end{subfloat}%
	\subfloat[][]{%{0.5/textwidth}
    \includegraphics[scale=0.35]{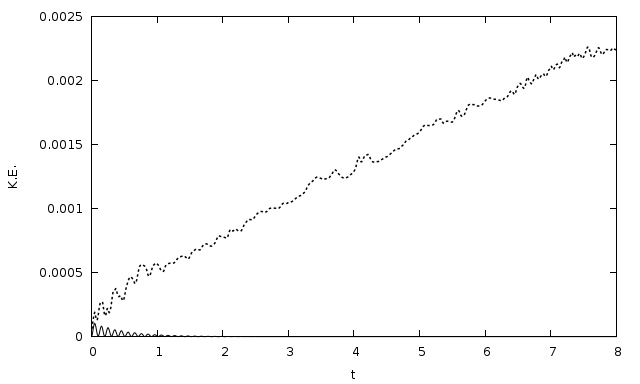}}
  %\end{subfloat}%
	\caption{Plot of TKE against time. The dotted lines are the results of using the conventional scheme while the solid lines are of using our new shceme. a) 2D. b) axis-symmetry.}
  \label{ocd_ke}
\end{figure}

\begin{figure}[h]
  \centering
	\subfloat[][]{%{0.5/textwidth}
    \includegraphics[scale=0.45]{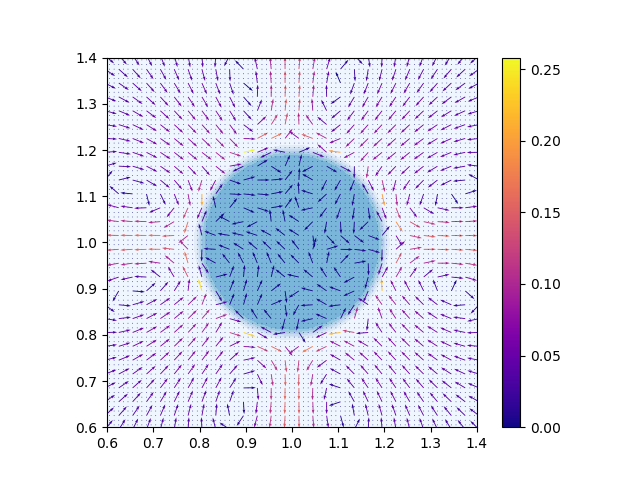}}
  %\end{subfloat}%
	\subfloat[][]{%{0.5/textwidth}
    \includegraphics[scale=0.45]{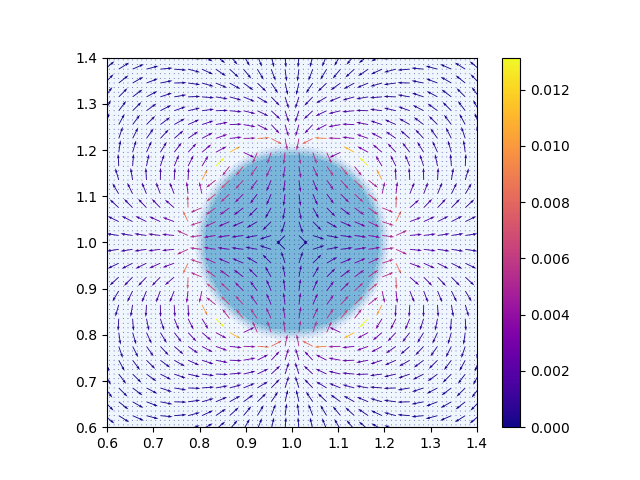}}
  %\end{subfloat}%
	\caption{Vector plot of velocity field of a 2D oscillating droplet at time $t=2$. a) Conventional scheme. b) New scheme.}
  \label{ocd_uv}
\end{figure}

Figure \ref{ocd_ke} shows the time series of total kinetic energy (TKE) for simulations with the grid 
size of $0.01$. We ran the simulations until time $t=8\, s$ and the total kinetic energy of fluid on a Cartesian grid is defined by: 
\begin{equation}
  \label{eqn:tke}
  TKE = \sum_{i,j} (u_{i,j}^2 + v_{i,j}^2) \Delta V_{i,j} 
\end{equation}
where $\Delta V_{i,j}$ is the cell's area (or volume in the axisymmetric case). The TKE is diminishing towards zero with the new scheme while it does not decrease properly for the conventional scheme due to the parasitic currents that persistently exist. Figure \ref{ocd_uv} further shows the velocity field of a 2D oscillating droplet at time $t=2\, s$. The parasitic currents are clearly visible when using the conventional scheme. Its magnitude is much larger than that of the physical flow field of the oscillating motion. As the parasitic currents are running in tangential direction to the interface, influence on the oscillating motion of the droplet is limited. However, fictituous kinetic energy due to the parasitic currents is persistently generated, and the total energy is not conserved. \par
\begin{figure}[h]
  \centering
	\subfloat[][]{%{0.5/textwidth}
    \includegraphics[scale=0.35]{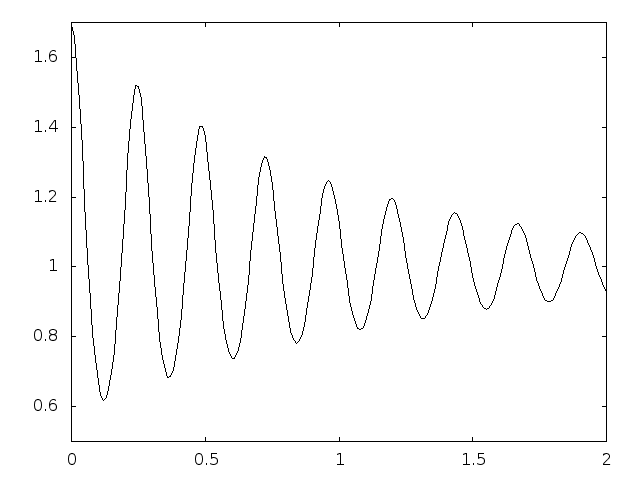}}
  %\end{subfloat}%
	\subfloat[][]{%{0.5/textwidth}
    \includegraphics[scale=0.35]{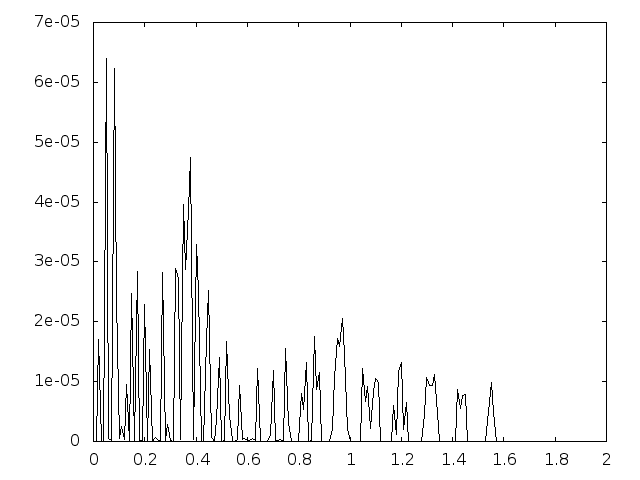}}
  %\end{subfloat}%
  \caption{Left: Plot of axis ratio against time. Right: Plot of $|\Sigma_{i,j} \vec{F}_{i,j}|/\Sigma_s |\vec{F}_{s}|$ where the numerator is absolute value of sum of surface tension force on Eulerian grid and denominator is sum of absolute force over all Lagrangian markers. }
  \label{ocd_deform}
\end{figure}

Finally, we have run a simulation where the droplet is initially deformed with a large amplitude. In this simulation, the $\beta$ in equation \eqref{eqn:osc} is set to $1.3$. The axis ratio and total absolute value of sum of surface tension force spread on the Eulerian grid over sum of absolute value on Lagrangian markers are shown in figure \ref{ocd_deform}. The maximum of total force error is at the order of $10^{-4}$ and it is too small to cause significant spurious movements of the droplet even under such a large deformation. \par

\begin{table}
\begin{center}
{\footnotesize
  \begin{tabular}{|c|c|c|c|c|c|}
   \hline
    & $\Delta x$ & Period (conventional) & discrepancy (conventional) & Period (new) & discrepancy (new)\\
   \hline
   2D & 0.02 & 0.2348 & 1.83$\%$ & 0.2411 & 4.57$\%$ \\
   \hline
   2D & 0.01 & 0.2334 & 1.23$\%$ & 0.2346 & 1.75$\%$ \\
   \hline
   2D & 0.005 & 0.2326 & 0.88$\%$ & 0.2328 & 0.97$\%$ \\
   \hline
   Axis-symmetry & 0.02 & 0.2044 & 2.53$\%$ & 0.2090 & 4.84$\%$ \\
   \hline
   Axis-symmetry & 0.01 & 0.2039 & 2.28$\%$ & 0.2024 & 1.53$\%$ \\
   \hline
  \end{tabular}
	}
  \caption{Oscillation periods (measured from the first cycle) of an oscillating droplet for different resolutions. The discrepancy is calculated by $|T_{sim} - T_{true}|/T_{true}\times 100\%$ where $T_{sim}$ is the period computed from the simulation result and $T_{true}$ is the theoretical value computed from eq. \eqref{eqn:osc_p} or \eqref{eqn:aosc_p}.}
  \label{table:osc}
\end{center}
\end{table}

\subsection{Free-fall droplet}
In this section, we apply our new scheme to the simulations of axisymmetric free-fall droplet with low density ratio and 2D free-fall water droplet through still air with large density ratio of $831.67$. In the first set of simulations, the droplet's shape and its Reynolds number at terminal velocity is compared to the experimental data for different Morton numbers and E\"{o}tvos numbers. In the second set, a circular water droplet of different diameters falls freely until it reaches its terminal velocity. The motivation of these simulations is to show that the new method is capable of dealing with simulations of long time-scale and large density ratio. \par
According to Burkingharm $\pi$ theorem, there are four dimensionless numbers governing the shape and terminal velocity of a free-fall droplet. These numbers are defined as: \\
\begin{equation}
  \label{eqn:non_dmn}
  \begin{split}
  d &= \frac{\rho_{droplet}}{\rho_{ambient}}, \\
  v &= \frac{\mu_{droplet}}{\mu_{ambient}}, \\
  \text{E\"{o}tvos number } &= g \frac{|\rho_{droplet} - \rho_{ambient}|D^2}{\sigma}, \\
  \text{Morton number } &= g \frac{\mu_{ambient}^{4}|\rho_{droplet} - \rho_{ambient}|D^2}
                         {\rho_{ambient}^{2} \sigma^{3}} 
  \end{split}
\end{equation}
where $d$ and $v$ are density and viscosity ratios respectively. 
Based on experimental data \cite{Cli}, there exists a relationship between the shape regime, Reynolds number, E\"{o}tvos number, and Morton number of a free-fall droplet or bubble with small density and viscosity ratios. Readers can refer to figure 2.5 in \cite{Cli} for the relationship. The shapes drawn in their figure are for bubbles. They should be upside-down for our simulations of a free-fall droplet. According to their figure, there are 6 different shape regimes depending on the Morton number and E\"{o}tvos number: Shperical, ellipsoidal, dimpled ellipsoidal-cap, skirted, wobbling, and spherical cap. We performed four free-fall droplet simulations with different sets of dimensionless numbers and compare the droplet's shape and Reynolds number with the data on their figure. These dimensionless numbers are tabulated in table \ref{table:afd_param}. For the first set of dimensionless numbers, the corresponding shape regime of a droplet at terminal velocity is spherical with no deformation. For the second, third, and fourth sets, the shapes are ellipsoidal, ellipsoidal with larger deformation, and dimpled ellipsoidal-cap respectively. These numbers are so chosen to demonstrate that stable simulations of a free-fall droplet with different magnitudes of deformation, from spherical shape to ellipsoidal-cap shape, at terminal velocity can be achieved by the new method. \par
\begin{table}
\begin{center}
  \begin{tabular}{|c|c|c|c|c|}
    \hline
		 & E\"{o}tvos number & Morton numer & d & v \\
		\hline
		1 & $0.3$ & $3.333 \times 10^{-4}$ & $10$ & $10$ \\
		\hline
		2 & $9$ & $1.440 \times 10^{-2}$ & $10$ & $10$ \\
		\hline
		3 & $10$ & $5.000 \times 10^{-4}$ & $10$ & $10$ \\
		\hline
		4 & $50$ & $100$ & $10$ & $10$ \\
		\hline
	\end{tabular}
	\caption{Four sets of parameters (E\"{o}tvos number, Morton numer, density ratio, and visocity ratio) for the simulations in this work.}
  \label{table:afd_param}
\end{center}
\end{table}

\begin{figure}[h]
  \centering
  \includegraphics[scale=0.6]{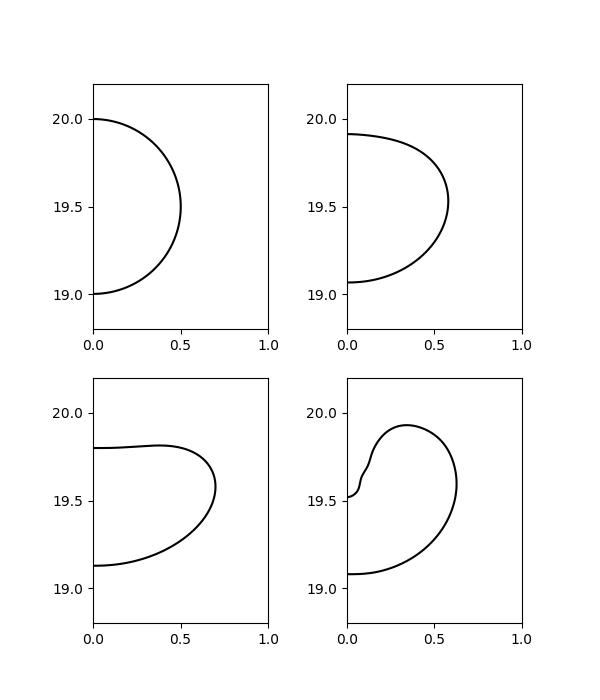}
  \caption{Snapshots of the droplets shape at terminal velocity for different parameter settings. Top-left: First set of dimensionless numbers. Top-right: Second set. Bottom-left: Third set. Bottom-right: Fourth set. The wavy interface in the bottom right figure near the axis of symmetry is caused by the wave motion traveling along the interface.}
  \label{afd_shape}
\end{figure}

\begin{figure}[h]
  \centering
  \includegraphics[scale=0.6]{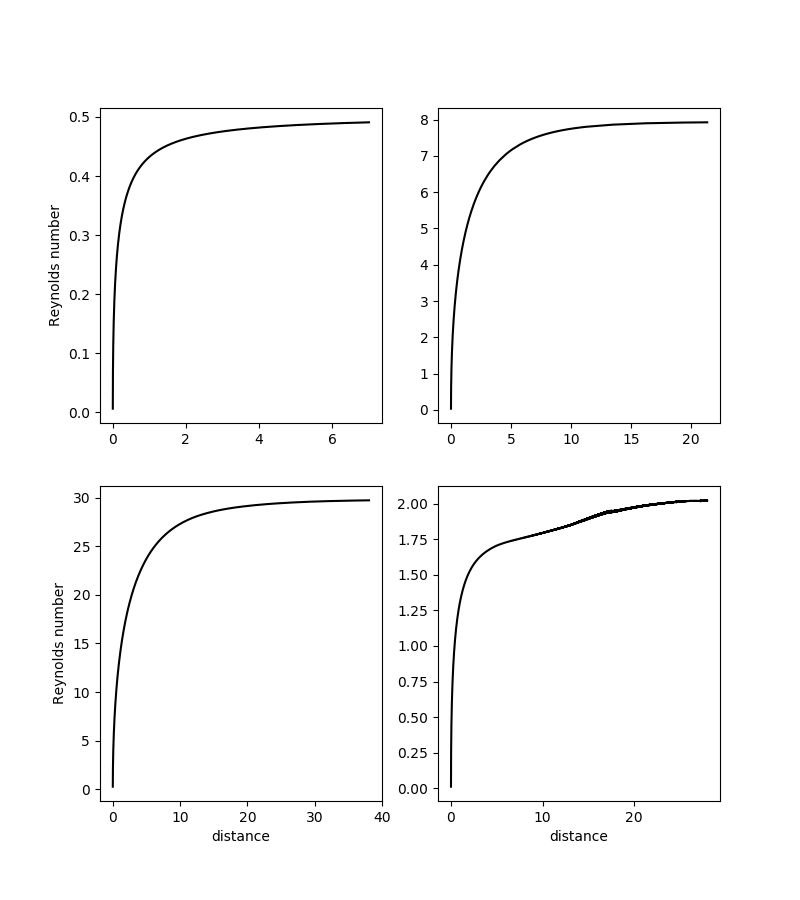}
  \caption{Plot of Reynolds number against dimensionless distance travelled for different parameter settings. Top-left: First set of dimensionless numbers. Top-right: Second set. Bottom-left: Third set. Bottom-right: Fourth set.}
  \label{afd_reynolds}
\end{figure}

The diameter is non-dimensionalized to one in the simulations. The grid size is $0.0125$ for the first two sets of 
parameters and $0.00625$ for the other sets. The droplet initially has a perfect spherical shape. The shape of the 
droplet for each set of parameters when it reaches its terminal velocity is shown in figure \ref{afd_shape}. Figure 
\ref{afd_reynolds} further shows the plot of time evolution of 
Reynolds number against nondimensionial distance that the droplet has travelled. The droplet's shape matches the corresponding shape regime provided in \cite{Cli} and the Reynolds number at terminal velocity is also in good agreement with their experimental data qualitatively. \par

\begin{figure}[h]
  \centering
  \includegraphics[scale=0.6]{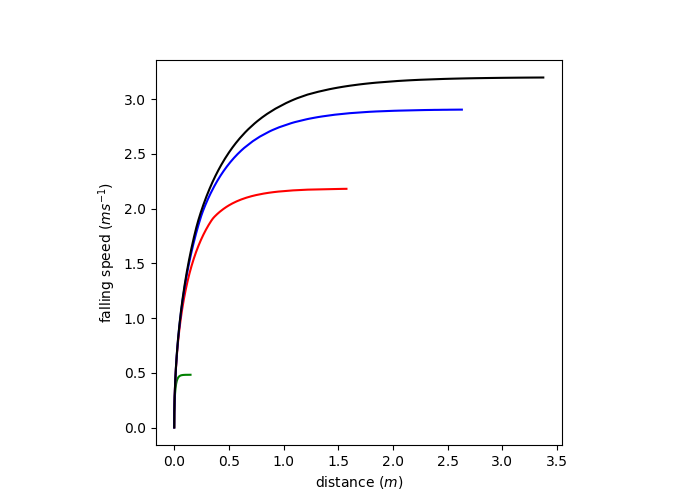}
  \caption{Plot of falling velocity (in meter per second) against distance travelled (in meter) for 2D water droplet of different diameters. Green: $0.1mm$. Red: $0.5mm$. Blue: $0.8mm$. Black: $1.0mm$.}
  \label{wfd_spd}
\end{figure}

\begin{figure}[h]
  \centering
  \subfloat[][]{%{0.5/textwidth}
    \includegraphics[scale=0.45]{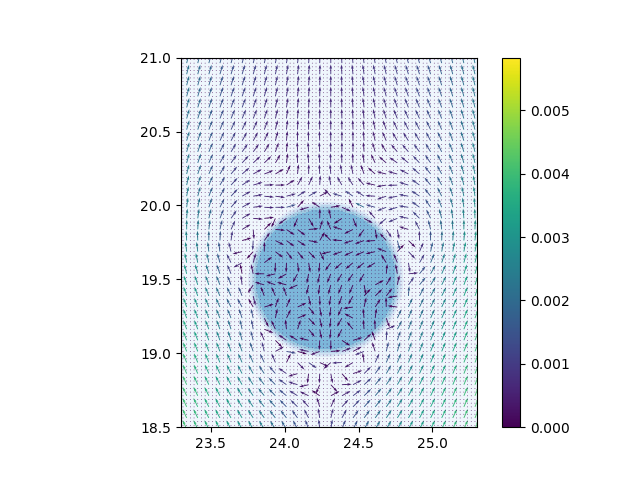}}
  %\end{subfloat}%
	\subfloat[][]{%{0.5/textwidth}
    \includegraphics[scale=0.45]{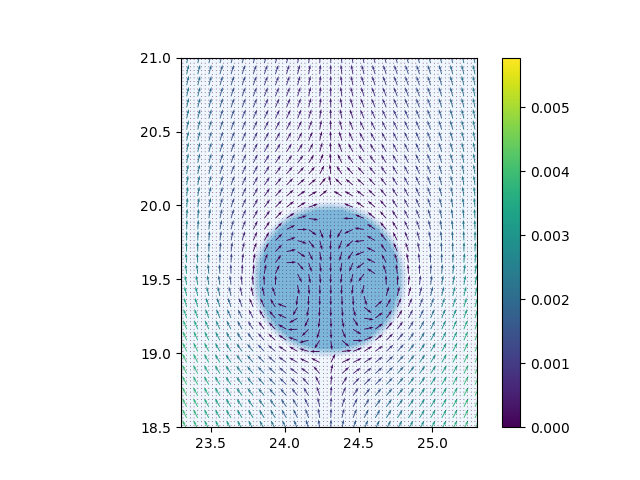}}
  %\end{subfloat}%
  \caption{Vector plot of velocity field of $0.1mm$ free-fall water droplet at terminal velocity. Left: Conventional scheme. Right: New scheme.}
  \label{wfd_incirc}
\end{figure}

In the simulations of 2D free-fall water droplet, we have chosen four different diameters $0.1\, mm$, 
$0.5\, mm$, $0.8\, mm$, $1.0\, mm$ to test the performance of the new scheme for high density ratio and 
surface tension force. The grid size is $0.0125$ for the simulation of droplet of diameter $1\, mm$ and $0.25$ for the others. A plot of falling velocity against distance travelled by the droplet is shown in figure \ref{wfd_spd}. The variation in velocity is less than $0.01$\% at the end of each simulation. This indicates that the numerical solutions converge properly to steady state solutions. From the figure, we also see that droplet with larger diameter travels longer before it reaches its terminal velocity. However, as no experimental 
data exist for free-fall 2D water droplets, no comparisons can be done to verify our numerical 
solutions. In figure \ref{wfd_incirc}, we plot the velocity field for $0.1\, mm$ water droplet at terminal velocity using the conventional IBM and our new IBM. The grid resolution is $0.0025\, mm$. The internal circulation is not visible with the conventional method due to the strong parasitic currents. The maximum magnitude of the parasitic currents is as large as $0.2\, ms^{-1}$ which is about $40$\% of the droplet's terminal velocity. Parasitic currents are weaker for a larger droplet because it has a smaller surface tension force. Further application of our new scheme to axisymmetric and 3D free-fall water droplet will be performed for comparisons with the experimental data and published in the near future. 

\newpage
\section{Conclusion}

In this work, we have shown that preserving the irrotational condition in the discretization of 
the delta function for the force spreading is the key to the elimination of parasitic currents. A new immersed boundary 
method with the discrete delta vector that satisfies the irrotational condition is developed. 
Moreover, we have proposed B-spline curve fitting of the Lagrangian markers by least squares. This enables 
smooth reconstruction of an interface at every time step to filter out unphysical fluctuations of 
curvature due to the spline interpolation that is very sensitive to locations of markers. 
The implementation of this new IBM has been verified through the static droplet and oscillating 
droplet test-cases. The results show that the parasitic currents are eliminated properly and the 
new method maintain a first order accuracy. The application of the irrotational delta vector is not limited to 
the IBM, but can be extended to other one-fluid models such as the VOF and the LS methods. 
Free-fall axisymmetric droplet and 2D water droplet simulations have subsequently been 
performed as an application of the new IBM. The results show good agreement with experimental data, 
demonstrating that the new method is capable of long-time-scale and high density ratio simulations. We hope that the new 
spreading operator that we proposed in this work can be a good alternative to the traditional delta function spreading operator 
when dealing with numerical problems where the internal or surface circulations and mixing are 
important. We are planning to extend it to a 2D interface for general 3D simulations of free-fall 
water droplet in the near future. \par

% produces the bibliography section when processed by BibTeX
\newpage
\bibliography{testbib}
\bibliographystyle{unsrt}

\newpage
\section{Figures}

\clearpage

\end{document}